\documentclass[a4paper,usenatbib,times,fleqn]{mn2e}
\usepackage{natbib,graphicx,amsmath,amsfonts,amssymb,times,txfonts,pstricks,lipsum}

\def\vg{\textbf{v}_{\mathrm{g}}}
\def\vd{\textbf{v}_{\mathrm{d}}}
\def\rhog{\rho_{\mathrm{g}}}
\def\rhod{\rho_{\mathrm{d}}}
\def\ts{t_{\mathrm{s}}}
\def\tp{t_{\mathrm{p}}}
\def\deltav{\Delta \textbf{v}}
\def\cs{c_{\mathrm{s}}}
\def\vb{\textbf{v}}
\def\vbz{\textbf{v}_{0}}
\def\rhogz{\rho_{\mathrm{g},0}}
\def\rhodz{\rho_{\mathrm{d},0}}
\def\rhoz{\rho_{0}}
\def\tstop{t_{\mathrm{stop}}}
\def\etheta{\mathbf{e}_{\theta}}
\def\ge{g_{\mathrm{e}}}
\def\vort{\mathbf{\omega}}
\def\hrho{\hat{\rho}}
\def\hrhod{\hat{\rho}_{\mathrm{d}}}
\def\hrhog{\hat{\rho}_{\mathrm{g}}}
\def\epsd{\epsilon}

\def\dst{\displaystyle}

\defcitealias{LP12a}{LP12a}
\defcitealias{LP12b}{LP12b}
\defcitealias{NSH86}{NSH86}

\title[Dusty gas with one fluid]{Dusty gas with one fluid}

\author[Laibe \& Price]{Guillaume Laibe$^{1,2}$, Daniel J. Price$^{1}$ \\
$^{1}$Monash Centre for Astrophysics and School of Mathematical Sciences, Monash University, Clayton, Vic 3800, Australia \\
$^{2}$School of Physics and Astronomy, University of St. Andrews, North Haugh, St. Andrews, Fife KY16 9SS, UK
}
\pagerange{\pageref{firstpage}--\pageref{lastpage}} \pubyear{2014}

\begin{document}
%
%
%


\def\jnl@style{\it}
\def\aaref@jnl#1{{\jnl@style#1}}

\def\aaref@jnl#1{{\jnl@style#1}}

\def\aj{\aaref@jnl{AJ}}                   
\def\araa{\aaref@jnl{ARA\&A}}             
\def\apj{\aaref@jnl{ApJ}}                 
\def\apjl{\aaref@jnl{ApJ}}                
\def\apjs{\aaref@jnl{ApJS}}               
\def\ao{\aaref@jnl{Appl.~Opt.}}           
\def\apss{\aaref@jnl{Ap\&SS}}             
\def\aap{\aaref@jnl{A\&A}}                
\def\aapr{\aaref@jnl{A\&A~Rev.}}          
\def\aaps{\aaref@jnl{A\&AS}}              
\def\azh{\aaref@jnl{AZh}}                 
\def\baas{\aaref@jnl{BAAS}}               
\def\jrasc{\aaref@jnl{JRASC}}             
\def\memras{\aaref@jnl{MmRAS}}            
\def\mnras{\aaref@jnl{MNRAS}}             
\def\pra{\aaref@jnl{Phys.~Rev.~A}}        
\def\prb{\aaref@jnl{Phys.~Rev.~B}}        
\def\prc{\aaref@jnl{Phys.~Rev.~C}}        
\def\prd{\aaref@jnl{Phys.~Rev.~D}}        
\def\pre{\aaref@jnl{Phys.~Rev.~E}}        
\def\prl{\aaref@jnl{Phys.~Rev.~Lett.}}    
\def\pasp{\aaref@jnl{PASP}}               
\def\pasj{\aaref@jnl{PASJ}}               
\def\qjras{\aaref@jnl{QJRAS}}             
\def\skytel{\aaref@jnl{S\&T}}             
\def\solphys{\aaref@jnl{Sol.~Phys.}}      
\def\sovast{\aaref@jnl{Soviet~Ast.}}      
\def\ssr{\aaref@jnl{Space~Sci.~Rev.}}     
\def\zap{\aaref@jnl{ZAp}}                 
\def\nat{\aaref@jnl{Nature}}              
\def\iaucirc{\aaref@jnl{IAU~Circ.}}       
\def\aplett{\aaref@jnl{Astrophys.~Lett.}} 
\def\apspr{\aaref@jnl{Astrophys.~Space~Phys.~Res.}}
\def\bain{\aaref@jnl{Bull.~Astron.~Inst.~Netherlands}} 
\def\fcp{\aaref@jnl{Fund.~Cosmic~Phys.}}  
\def\gca{\aaref@jnl{Geochim.~Cosmochim.~Acta}}   
\def\grl{\aaref@jnl{Geophys.~Res.~Lett.}} 
\def\jcp{\aaref@jnl{J.~Chem.~Phys.}}      
\def\jgr{\aaref@jnl{J.~Geophys.~Res.}}    
\def\jqsrt{\aaref@jnl{J.~Quant.~Spec.~Radiat.~Transf.}}
\def\memsai{\aaref@jnl{Mem.~Soc.~Astron.~Italiana}}
\def\nphysa{\aaref@jnl{Nucl.~Phys.~A}}   
\def\physrep{\aaref@jnl{Phys.~Rep.}}   
\def\physscr{\aaref@jnl{Phys.~Scr}}   
\def\planss{\aaref@jnl{Planet.~Space~Sci.}}   
\def\procspie{\aaref@jnl{Proc.~SPIE}}   

\let\astap=\aap
\let\apjlett=\apjl
\let\apjsupp=\apjs
\let\applopt=\ao

\label{firstpage}
\bibliographystyle{mn2e}
\maketitle

\begin{abstract}
In this paper, we show how the two-fluid equations describing the evolution of a dust and gas mixture can be re-formulated to describe a single fluid moving with the barycentric velocity of the mixture. This leads to evolution equations for the total density, momentum, the differential velocity between the dust and the gas phases and either the dust-to-gas ratio or the dust fraction. The equations are similar to the usual equations of gas dynamics, providing a convenient way to extend existing codes to simulate two-fluid mixtures without modifying the code architecture.

 Our approach avoids the inherent difficulties related to the standard approach where the two phases are separate and coupled via a drag term. In particular, the requirements of infinite spatial and temporal resolution as the stopping time tends to zero are no longer necessary. This means that both small and large grains can be straightforwardly treated with the same method, with no need for complicated implicit schemes. Since there is only one resolution scale the method also avoids the problem of unphysical trapping of one fluid (e.g. dust) below the resolution of the other. 
 
 We also derive a simplified set of equations applicable to the case of strong drag/small grains, consisting of the standard fluid equations with a modified sound speed, plus an advection-diffusion equation for the dust-to-gas ratio. This provides a simple and fast way to evolve the mixture when the stopping time is smaller than the Courant timestep. We present a Smoothed Particle Hydrodynamics implementation in a companion paper.

\end{abstract}


\section{Introduction}
\label{sec:intro}
Dust. It pervades the interstellar medium, molecular clouds, young stellar systems and protoplanetary discs, and is the material from which planets are formed. Grain properties are inferred via a range observational techniques at multiple wavelengths, including spectral emission distributions, scattered light images, thermal emission maps, mid-infrared spectroscopy, polarimetry and molecular emissions (see \citealt{Pinte2008} for a detailed analysis). We can thus determine the spatial distribution, composition, crystallinity and even more importantly, grain sizes $s$, The largest fraction of the dust mass lies in the small grains since the dust size distribution is found to vary as $n\left( s \right) \simeq s^{-3.5}$ \citep{Mathis1977}.

Simulating grain evolution in astrophysical objects is of primary importance since they play a crucial role in transferring energy, being the main source of thermal infrared emission. They thus affect the thermodynamics of the gas and --- in turn --- its dynamics. This is key to astrophysical processes that are controlled by the ability of the system to dissipate its gravitational energy, such as star and planet formation. However, computing the local opacities self-consistently taking into account the local evolution of the dust population and in general, simulating dust and gas mixtures evolution is quite challenging. Dust grains mainly interact with the surrounding gas by microscopic collisions between gas molecules and the dust grains. This results in a macroscopic drag force which damps the differential velocity between the two phases. The typical timescale for the damping is called the stopping time $t_{\rm{s}}$, which usually increases when the grain size decreases. Numerical methods thus need to be efficient over a wide range of dust sizes encountered in astrophysical systems. This is particularly the case in protoplanetary discs, where grains with different sizes become spatially differentiated by dynamical processes \citep{Weidendust1977,NSH86,GaraudLin2004,Youdin2005,Carballido2006}.

In two previous papers (\citealt{LP12a,LP12b}, hereafter \citetalias{LP12a,LP12b}), we have detailed and benchmarked a Lagrangian Smoothed Particle Hydrodynamics (SPH) formalism to treat astrophysical dust-gas mixtures. This scheme used a variable SPH smoothing length, SPH terms for the conservative part of the equations derived from a Lagrangian, and a `double-hump' shaped kernel that was found to improve the accuracy of the drag terms in both linear and non linear drag regimes. However, alongside the numerous tests we performed, we faced two rather fundamental difficulties which were not related to SPH but inherent to the two fluid formalism. Firstly, if dust particles concentrate below the resolution length of the gas phase, they tend to become trapped there, since they no longer feel any differential forces from the gas (see \citealt{Ayliffe2012} for a discussion of this issue in SPH simulations; and \citealt{PF10} for the same issue in a grid-based context). Secondly, when handling the specific problem of strong drag regimes (corresponding to small grains), we found two limitations which lead to a prohibitive computational cost: 1) the drastically small time steps required for the numerical stability of explicit schemes or the complexity of the implicit schemes involved, and less trivially 2) a high spatial resolution required to resolve the differential velocity between the gas and the dust in order to simulate the correct physical dissipation rate. The latter occurs because even if the differential velocity between the fluids is damped after a few stopping times $t_{\rm s}$, the gas pressure causes a small spatial dephasing between the gas and dust. When the resolution is too low ($\Delta x \lesssim t_{\rm s} c_{\rm s}$, see \citetalias{LP12a}), the dephasing from numerical simulations is artificially too large and the energy is over-dissipated.

This means that it is essentially impossible to simulate small grains accurately using the two fluid approach with standard fluids codes, since both infinitely small time steps and an infinitely large spatial resolution are required in the limit $t_{\rm s}\to 0$. Worse still, this limit corresponds to the rather obvious limit in which the two fluids are perfectly coupled and move precisely as a single fluid, albeit with a sound speed modified by the dust to gas ratio. For astrophysics this means that it is not currently possible to simulate the small, micron to cm sized grains accurately with \emph{any} existing dust-gas code where dust is simulated using particles, and certainly not possible to simulate both small and large (metre-to-planetesimal sized) grains with the same technique.

In this paper, we show how the equations describing gas-dust mixtures can be reformulated to represent a single fluid moving with the barycentre of the mixture, leading to a set of equations only slightly modified from the usual equations of gas dynamics, with additional evolution equations for the differential velocity and the dust-to-gas ratio. This approach, though initially developed with the large drag/small grain regime in mind, turns out to be both general and elegant, since the important physical quantities of the mixture are computed directly, avoiding all of the artificial complications which arise in the two-fluid treatment.

 The equations for the evolution of the single fluid are derived in Sec.~\ref{sec:single} in primitive and conservative forms for the formalism to be relevant for both Lagrangian and Eulerian methods. This set of equations is completely general and can be used to simulate both large and small grains. In Sec.~\ref{sec:approx}, we show how they can be further simplified in the specific limit of small grains and subsonic differential motion, leading to the standard equations of gas dynamics (with a modified sound speed), coupled with an advection-diffusion equation for the dust-to-gas ratio. In Sec.~\ref{sec:tests} we demonstrate that the main physical effects associated with dust, including linear waves, shocks and the streaming instability, can all be captured with this simplified approach, and give the appropriate criterion for the use of the simplified formulation in numerical codes.

\section{Single fluid model}
\label{sec:single}

\subsection{Two fluid equations}
\label{sec:general}

In astrophysical problems, dust and gas mixtures are usually treated by two continuous phases that interact via a drag term (\citealt{Saffman1962}; see e.g. \citetalias{LP12a} for a particular implementation). The dust fluid is treated as a pressureless fluid. The equations for the conservation of density and momentum are therefore given by
\begin{eqnarray}
\frac{\partial \rhog}{\partial t} + \nabla\cdot\left ( \rhog \vg \right) & = & 0 \label{eq:mass_gas},\\
\frac{\partial \rhod}{\partial t} + \nabla\cdot\left ( \rhod \vd \right) & = & 0 \label{eq:mass_dust}, \\
\rhog \left( \frac{\partial \vg}{\partial t} + \vg\cdot\nabla \vg \right)  & = & \rhog \textbf{f} +  K  (\vd - \vg)  - \phantom{.}\nabla P_{\rm g} \label{eq:momentum_gas},\\[1em]
\rhod \left( \frac{\partial \vd}{\partial t} + \vd\cdot\nabla \vd \right) & = &  \rhod \textbf{f} -  K  (\vd - \vg) \label{eq:momentum_dust},
\end{eqnarray}
where the subscripts ${\rm g}$ and ${\rm d}$ refer to the gas and dust, respectively, and $K$ is the drag coefficient which is a function of the local gas and dust parameters, as well as the differential velocity between the fluids (see \citetalias{LP12b} for an extensive discussion of drag regimes). In the following, we will denote $\cs$ the gas sound speed such as $\delta  P_{\rm g} = \cs^{2} \delta \rhog $ and $\ts$, the typical drag stopping time given by
\begin{equation}
\ts \equiv \frac{\rhod \rhog}{K\left(\rhog + \rhod \right)} .
\label{eq:ts}
\end{equation}
Some studies adopt $t_{\mathrm{stop}} = \rhod/K$ for the stopping time. We use the definition given by Eq.~\ref{eq:ts} since it is more physically relevant as we will see hereafter. Qualitatively, two limiting behaviours occur for the mixture's evolution, depending on the value of $\ts$ compared to the other physical typical time. If $\ts$ is large (weak drag, i.e. large grains in astrophysics), the drag dissipates the differential kinetic energy between the phases slowly and is essentially perturbative. From a numerical point of view, such drag terms can be integrated by a straightforward explicit integration. If $\ts$ is small (strong drag, i.e. small grains in astrophysics), the drag controls the evolution of the mixture since momentum between the two phases is almost instantaneously exchanged. The behaviour of the mixture becomes less intuitive. In \citetalias{LP12a}, we have illustrated the behaviour of a gas and dust mixture at strong drag regimes with the \textsc{dustywave} problem. After a typical time $\ts$, the initial differential velocity between the fluids is damped and the barycentre of the fluid propagates with a modified sound speed $\tilde{\cs}$ (see below,  Eq.~\ref{eq:cstilde}). However, since $\ts$ remains finite, the gas pressure makes the gas propagate a small distance $\cs \ts$ with respect to the dust and both waves in the gas and the dust are slightly dephased. This small dephasing is then damped by the drag but regenerated by the pressure, leading to a dissipation in the evolution of the mixture, while both phases remain closely coupled.

 The evolution equation for the specific internal energy of the gas is given by
\begin{equation}
\frac{\partial u}{\partial t} + (\vg \cdot \nabla) u = -\frac{P_{\rm g}}{\rho_{g}} (\nabla\cdot \vg) + K (\vd - \vg)^{2}, \label{eq:dudt}
\end{equation}
the last term representing the dissipation of heat due to drag.

\subsection{One-fluid model}
\label{sec:onefluid}
Without loss of generality, Eqs.~\ref{eq:mass_gas} -- \ref{eq:momentum_dust} and \ref{eq:dudt} can be reformulated as a single fluid, moving with the barycentric velocity,
\begin{eqnarray}
\vb & \equiv & \frac{\rhog \vg + \rhod \vd}{\rhog + \rhod} \label{eq:def_vb},
\end{eqnarray}
and evolving the differential velocity between the two phases, $\deltav$, defined according to
\begin{eqnarray}
\Delta {\bf v} & \equiv & \vd -\vg  \label{eq:def_deltav}.
\end{eqnarray}
In the barycentric frame, the total density $\rho \equiv \rhog + \rhod$ and the dust to gas ratio $\rhod / \rhog$ are the natural quantities to study the evolution of the mixture. Using the identities
\begin{eqnarray}
\vg & = & \vb - \frac{\rhod}{\rho} \deltav \label{eq:newvg},\\
\vd & = & \vb + \frac{\rhog}{\rho} \deltav \label{eq:newvd},
\end{eqnarray}
Eqs.~\ref{eq:mass_gas} -- \ref{eq:momentum_dust} become
\begin{eqnarray}
\frac{\partial \rho}{\partial t} + \nabla\cdot\left( \rho \vb \right) & = & 0  \label{eq:mass_rho_euler},\\
\frac{\partial \vb}{\partial t} + (\vb\cdot\nabla) \vb & = & \mathbf{f} - \frac{\nabla P_{\mathrm{g}}}{\rho} - \frac{1}{\rho}\nabla\cdot \left(\frac{\rhog \rhod}{\rho} \deltav \deltav \right)   \label{eq:momentum_bary_euler},\\[1em]
\frac{\partial}{\partial t} \left(\frac{\rhod}{\rhog} \right) + \vb\cdot\nabla \left(\frac{\rhod}{\rhog}\right) & = & -\frac{\rho}{\rhog^{2}} \nabla \cdot \left(\frac{\rhog \rhod}{\rho} \deltav \right) \label{eq:dtgevol_euler} , \\
\frac{\partial \deltav}{\partial t} + (\vb \cdot \nabla) \deltav  & = &  - \frac{\deltav}{\ts} + \frac{\nabla P_{\mathrm{g}}}{\rhog} \nonumber \\
&& - (\deltav \cdot \nabla) \vb + \frac{1}{2}\nabla \left( \frac{\rhod - \rhog}{\rhod + \rhog} \deltav ^{2} \right)  \label{eq:momentum_deltav_euler}.
\end{eqnarray}
The evolution of the gas internal energy becomes
\begin{eqnarray}
\frac{\partial u}{\partial t} + (\vb \cdot \nabla) u& = & -\frac{P_{\mathrm{g}}}{\rhog} (\nabla\cdot\vb_{g}) + \frac{\rhod}{\rho}\left( \deltav\cdot\nabla\right) u + \frac{\rhod}{\rho} \frac{\deltav^{2}}{\ts},
\end{eqnarray}
or equivalently, the entropy evolves according to
\begin{eqnarray}
T \frac{\partial s}{\partial t} + (\vb \cdot \nabla) s & = &  \frac{\rhod}{\rho}\left( \deltav\cdot\nabla\right) s + \frac{\rhod}{\rho} \frac{\deltav^{2}}{\ts} , \label{eq:news}
\end{eqnarray}
where $T$ is the local gas temperature. As expected, the differential velocity between the gas and the dust is a dissipative and irreversible source of entropy.

 In the Lagrangian frame comoving with the fluid barycentre, the equations can be simplified further using the total time derivative
\begin{eqnarray}
\frac{\mathrm{d}}{\mathrm{d}t} = \frac{\partial}{\partial t} +  \vb\cdot\nabla,
\label{eq:defddt}
\end{eqnarray}
such that the evolution of the position, $\textbf{X}$, of a fluid particle of this single fluid is given by ${\mathrm{d} \textbf{X} }/{\mathrm{d}t} = \vb$. Thus Eqs.~\ref{eq:mass_rho_euler}--\ref{eq:momentum_deltav_euler} simplify to
\begin{eqnarray}
\frac{{\rm d} \rho}{{\rm d} t}& = & - \rho (\nabla\cdot\vb), \label{eq:mass_rho} \\
\frac{{\rm d} \vb}{{\rm d} t} & = & \mathbf{f} - \frac{\nabla P_{\mathrm{g}}}{\rho} - \frac{1}{\rho}\nabla\cdot \left(\frac{\rhog \rhod}{\rho} \deltav \deltav \right)   \label{eq:momentum_bary},\\[1em]
\frac{{\rm d}}{{\rm d} t} \left(\frac{\rhod}{\rhog} \right) & = & -\frac{\rho}{\rhog^{2}} \nabla \cdot \left(\frac{\rhog \rhod}{\rho} \deltav \right) \label{eq:dtgevol} , \\
\frac{{\rm d} \deltav}{{\rm d} t}  & = &  - \frac{\deltav}{\ts} + \frac{\nabla P_{\mathrm{g}}}{\rhog} - (\deltav \cdot \nabla) \vb + \frac{1}{2}\nabla \left( \frac{\rhod - \rhog}{\rhod + \rhog} \deltav ^{2} \right), \label{eq:momentum_deltav}
\end{eqnarray}
while the internal energy equation is given by
\begin{equation}
\frac{{\rm d} u}{{\rm d} t} =  -\frac{P_{\mathrm{g}}}{\rhog} (\nabla\cdot\vg) +   \frac{\rhod}{\rho}\left( \deltav\cdot\nabla\right) u  + \frac{\rhod}{\rho} \frac{\deltav^{2}}{\ts}. \label{eq:newu}
\end{equation}
The specific entropy of the gas evolves according to
\begin{equation}
\frac{{\rm d} s}{{\rm d} t} =  \frac{\rhod}{T \rho} \frac{\deltav^{2}}{\ts} ,
\end{equation}
showing that the drag is the only source of entropy in the mixture.

 Throughout this section, we have assumed that the volume occupied by the dust grains is negligible. For astrophysical applications --- with micron to kilometre-sized grains in simulations on AU or parsec scales --- this is an extremely good approximation, but it can be important in non-astrophysical problems (see \citet{FanZhu} for various examples). For completeness we give the one fluid equations generalised to finite volume grains in Appendix~\ref{sec:theta}.
 
 It should be noted that while physical, the use of the dust-to-gas ratio introduces an artificial singularity in the equations when the mixture is only made of dust $(\rhog = 0)$. A convenient way to overcome this difficulty is to use the dust fraction $\epsd =\rhod / \rho$ instead of the dust-to-gas ratio. The gas and the dust densities are calculated according to $\rhog = \left( 1 - \epsd \right) \rho$ and $\rhod = \epsd \rho$ respectively. Eqs.~\ref{eq:mass_rho} -- \ref{eq:newu} become:

\begin{eqnarray}
\frac{{\rm d} \rho}{{\rm d} t}& = & - \rho (\nabla\cdot\vb), \label{eq:epsd_rho} \\
\frac{{\rm d} \epsd}{{\rm d} t}& = & -\frac{1}{\rho} \nabla \cdot \left[ \epsd\left(1 - \epsd \right) \rho \deltav \right] \label{eq:epsd_evol} , \\
\frac{{\rm d} \vb}{{\rm d} t} & = &  - \frac{\nabla P_{\mathrm{g}}}{\rho}  - \frac{1}{\rho}\nabla\cdot \left[ \epsd\left(1 - \epsd \right) \rho \deltav \deltav \right] + \mathbf{f}\label{eq:epsd_momentum},\\[1em]
\frac{{\rm d} \deltav}{{\rm d} t}  & = &  - \frac{\deltav}{\ts} + \frac{\nabla P_{\mathrm{g}}}{\left(1 - \epsilon \right)\rho} - (\deltav \cdot \nabla) \vb + \frac{1}{2}\nabla \left[ \left(2 \epsd - 1 \right) \deltav ^{2} \right], \label{eq:epsd_deltav} \\
\frac{{\rm d} u}{{\rm d} t} & = &  -\frac{ P_{\mathrm{g}}}{\left(1 - \epsd \right) \rho} \nabla\cdot\left( \vb - \epsd \deltav \right) +   \epsd \left( \deltav\cdot\nabla\right) u  + \epsd  \frac{\deltav^{2}}{\ts}, \label{eq:epsd_u}
\end{eqnarray}
where the stopping time $\ts$ reads
\begin{equation}
\ts = \frac{ \epsd\left(1 - \epsd \right) \rho}{K } .
\label{eq:def_ts}
\end{equation}

\subsection{Advantages of the one fluid approach}
While mathematically equivalent to Eqs.~\ref{eq:mass_gas} -- \ref{eq:dudt}, the barycentric formulation of the dusty gas equations has a number of key advantages for the numerical solution of dust-gas mixtures. In particular:
\begin{enumerate}
\item The equations can be solved on a single fluid that moves with the barycentric velocity $\vb$, rather than requiring two fluids. In turn, this implies only one resolution scale in numerical models, avoiding the problems associated with mismatched spatial resolutions discussed above \citep[c.f.][]{PF10,Ayliffe2012,LP12a}.
\item The form of the continuity and acceleration equations (Eqs.~\ref{eq:mass_rho} and \ref{eq:momentum_bary}) are similar or identical to the usual equations of hydrodynamics, with a minor modification to the pressure gradient and one additional term in the acceleration equation.
\item The dust-to-gas ratio, the critical parameter in most astrophysical problems, is explicitly evolved.  Furthermore, both the physics producing a change in the dust-to-gas ratio, and the limit in which the dust-to-gas ratio is constant, are clear.
\item Drag terms between the two fluids do not have to be explicitly evaluated, meaning treatment of complicated or non-linear drag regimes is straightforward.
\item The evolution equation for $\deltav$ (Eq.~\ref{eq:momentum_deltav}) is analogous to the induction equation for magnetohydrodynamics or the evolution of vorticity in incompressible flows, with additional source ($\nabla P_{\rm g}/\rhog$) and decay ($-\deltav/\ts$) terms.
\item Implicit treatment of the decay term in Eq.~\ref{eq:momentum_deltav} in the limit of $\ts \to 0$ can be trivially achieved using operator splitting, since the exact solution is known.
\item The equations can be simplified further in the limit of strong drag/short stopping times, as we discuss below.
\end{enumerate}

 Eqs.~\ref{eq:mass_rho}, \ref{eq:momentum_bary} and \ref{eq:momentum_deltav} have been used in a reduced form (assuming an incompressible fluid) for analytic studies of instabilities in protoplanetary discs \citep{Youdin2005,Chiang2008,Barranco2009,Lee2010,Jacquet2011}. However, Eq.~\ref{eq:dtgevol} (or equivalently Eq.~\ref{eq:epsd_evol}) --- the most important equation in the barycentric formulation --- has to our knowledge not been derived elsewhere.

\subsection{Physical interpretation}
\label{sec:inter}
Eq.~\ref{eq:mass_rho} is a standard equation of mass conservation for the total mass of the system. Eq.~\ref{eq:momentum_bary} is also similar to a single fluid momentum conservation equation, except that 1) the gas pressure gradient is divided by the total density of the fluid, thus taking into account the inertia of the dust and 2) the dissipated energy from the differential velocity between the fluid acts like a kinetic pressure for the fluid. Indeed, in Eq.~\ref{eq:newe} the term $\frac12 \rho \vb^{2}$ is the dynamical kinetic energy of the mixture, while the second term is the density of energy internal to the mixture (which is equivalent to a pressure). The effect of these terms on the evolution of the fluid vorticity $\vort = \nabla \times \vb$ is given by the relation
\begin{eqnarray}
\frac{\partial \vort}{\partial t} + (\vb \cdot \nabla) \vort & = & \left( \vort\cdot\nabla\right) \vb - \vort (\nabla\cdot\vb) + \nabla \times \mathbf{f} \nonumber\\
&& + \frac{1}{\rho^{2}}\nabla{\rho} \times \nabla\left( P_{\mathrm{g}} +\frac{\rhog\rhod}{\rho}  \deltav^{2}  \right)\cdot\label{eq:vort}
\end{eqnarray}
The first terms of Eq.~\ref{eq:vort} are similar to the vorticity equation for a single gaseous fluid. However, vorticity can also be created from the last term of Eq.~\ref{eq:vort} which is specific to the mixture. First, even if the gas is a barotopic fluid, the mixture is usually not barotropic since $ \nabla P_{\mathrm{g}} $ is in general not colinear with $\nabla \rhod$. Second, the kinematic pressure is an additional source of vorticity for the fluid. It should finally be noted that the total helicity of the fluid \citep{Moffat1992} is conserved but the local helicity flux has to be modified by adding the dynamical pressure as well.

Eq.~\ref{eq:dtgevol} shows that in absence of any differential velocity between the fluids, the dust to gas ratio is advected with the mixture. This equation can alternatively be rewritten
\begin{eqnarray}
\frac{\partial}{\partial t} \left( \frac{\rhod}{\rhog}  \right) + (\vb \cdot \nabla)  \left( \frac{\rhod}{\rhog}  \right) & = & - \frac{\rhod}{\rhog} (\nabla\cdot\deltav) - \frac{\rhod}{\rhog} \frac{1}{\rho} \left( \deltav\cdot\nabla \right) \rho  \nonumber \\
&& - \frac{1 - \rhod / \rhog}{1 + \rhod/\rhog}\left( \deltav\cdot\nabla \right)   \left( \frac{\rhod}{\rhog}  \right)\cdot\label{eq:dtgevolbis}
\end{eqnarray}
Interestingly, it becomes transparent that even if $\deltav$ and $\rho$ are constant, the dust-to-gas ratio can change if $\rhod$ and $\rhog$ are different (this occurs in the case of the streaming instability).

Eq.~\ref{eq:momentum_deltav} shows that the source of differential velocity between the fluids is the pressure gradient (i.e., if $\nabla P_{\mathrm{g}} = 0$ and $\deltav = 0$ initially, no differential velocity is generated). This differential velocity is also damped by the drag term. Eq.~\ref{eq:momentum_deltav} is the only equation where $\ts$ is involved.

It should be noted that the total energy of the mixture $E_{\rm m}$ is
\begin{equation}
E_{\rm m} =   m\left( \frac{1}{2}  \vb^{2} + u_{\rm m} \right),
\end{equation}
where
\begin{equation}
 u_{\rm m} = \left(1 - \epsilon \right)u(s,(1 - \epsilon)\rho) + \frac{1}{2} \epsd \left(1 - \epsd \right) \deltav^{2} ,
\end{equation}
$u$ and $s$ denoting the internal energy and entropy of the gas (and not of the mixture) respectively. Thus, the kinetic energy of the mixture is not the sum of the kinetic energies of its components since a term arising from the differential velocities goes into the internal energy.

\subsection{Generalised formulation with arbitrary gas/dust forces}
\label{sec:gen}
 In general there may be additional forces that act separately on either the gas phase or the dust phase beyond those given in Eqs.~\ref{eq:mass_gas}--\ref{eq:momentum_dust}. Examples include viscous and magnetohydrodynamic forces. In the most general case, the one fluid equations are given by
\begin{eqnarray}
\frac{{\rm d} \rho}{{\rm d} t}& = & - \rho (\nabla\cdot\vb), \label{eq:genmass_rho} \\
\frac{{\rm d} \vb}{{\rm d} t} & = & \frac{\rhog}{\rho} \mathbf{f}_{\rm g} + \frac{\rhod}{\rho} \mathbf{f}_{\rm d} - \frac{1}{\rho}\nabla\cdot \left(\frac{\rhog \rhod}{\rho} \deltav \deltav \right) + \mathbf{f}\label{eq:genmomentum_bary},\\[1em]
\frac{{\rm d}}{{\rm d} t} \left(\frac{\rhod}{\rhog} \right) & = & -\frac{\rho}{\rhog^{2}} \nabla \cdot \left(\frac{\rhog \rhod}{\rho} \deltav \right) \label{eq:gendtgevol} , \\
\frac{{\rm d} \deltav}{{\rm d} t}  & = &  - \frac{\deltav}{\ts} + ({\bf f}_{\rm d} - {\bf f}_{\rm g}) - (\deltav \cdot \nabla) \vb + \frac{1}{2}\nabla \left( \frac{\rhod - \rhog}{\rhod + \rhog} \deltav ^{2} \right), \label{eq:genmomentum_deltav}
\end{eqnarray}
where ${\bf f}_{\rm g}$ refers to the forces acting only on the gas phase (e.g. pressure and viscous forces), ${\bf f}_{\rm d}$ is any force acting only on the dust phase, and ${\bf f}$ is the force acting on both phases (e.g. gravity). As previously, Eq.~\ref{eq:epsd_evol} can equivalently be used in place of Eq.~\ref{eq:gendtgevol}.

\subsection{Conservative formulation}
An important constraint on any numerical implementation is that all of the conservation laws should be satisfied. In Paper~II we show that our SPH implementation of Eqs.~\ref{eq:epsd_rho}--\ref{eq:def_ts} conserves all of these quantities exactly, demonstrating that there is no disadvantage with respect to the two-fluid formulation in terms of conservation properties. For Eulerian codes this may be demonstrated by showing that the equations can be written as a hyperbolic system in conservative form.

The conservative part of the local equations of evolution of the mixture can be derived directly from the conservation of physical quantities over a volume of fluid $V$. This volume moves with a velocity $\mathbf{U}$, which can be the fluid velocity $\mathbf{v}$ (for the mixture, the barycentric velocity) or a different velocity. Denoting $\mathrm{d}S\mathbf{n}$ the elementary surface vector of the volume, the transport theorem provides the evolution of an integral quantity over $V$:
\begin{equation}
\frac{\delta}{\delta t} \left( \int_{V} b \mathrm{d}V \right) = \int_{V} \frac{\partial b}{\partial t} \mathrm{d}V + \int_{S} b\left(\mathbf{U}\cdot\mathbf{n} \right)\mathrm{d}S,
\label{eq:transport}
\end{equation}
where $b$ is a tensorial field of any order. Additional discontinuities in the quantity $b$ over the volume $V$ would add additional terms in the right-hand side of Eq.~\ref{eq:transport}, which we have neglected. If $\mathbf{U} = \mathbf{v}$, then $\frac{\delta}{\delta t} = \frac{\mathbf{d}}{\mathbf{d}t}$. 

\subsubsection{Conservation of mass}
The mass of gas and dust contained in the volume $V$ are given by
\begin{eqnarray}
M_{\rm g} & \equiv &  \int_{V} \rhog \mathrm{d}V =  \int_{V} \left(1 - \epsilon \right)\rho \mathrm{d}V, \label{eq:def_mg}\\
M_{\rm d} & \equiv &  \int_{V} \rhod \mathrm{d}V =  \int_{V}  \epsilon \rho \mathrm{d}V, \label{eq:def_md} 
\end{eqnarray}
From Eq.~\ref{eq:transport} we obtain
\begin{eqnarray}
\frac{\mathrm{d}_{\rm g} M_{\rm g}}{\mathrm{d} t} &  = & 0  , \label{eq:int_mg} \\
\frac{\mathrm{d}_{\rm d} M_{\rm d}}{\mathrm{d} t} &  = & 0  , \label{eq:int_md}
\end{eqnarray}
where $\dst \frac{\mathrm{d}_{\rm g}}{\mathrm{d} t}  = \dst  \frac{\partial}{\partial t} + \mathbf{v}_{\rm g}\cdot\nabla$ and $\dst  \frac{\mathrm{d}_{\rm d}}{\mathrm{d} t}  =\dst  \frac{\partial}{\partial t} + \mathbf{v}_{\rm d}\cdot\nabla$ are the comoving derivatives for the gas and the dust, respectively. Using the transport theorem with $\mathbf{U} = \vg$ and $\mathbf{U} = \vd$, respectively, and the divergence theorem, Eqs.~\ref{eq:int_mg} -- \ref{eq:int_md} result in two integral equations whose integrands are zero, implying two local conservation equations:
\begin{eqnarray}
\frac{\partial \rho \left(1 -  \epsilon\right)}{\partial t} + \nabla\cdot\left[ \rho\left( 1 - \epsilon\right) \vb - \rho \epsilon \left(1 - \epsilon \right) \deltav \right] & = & 0 \label{eq:cons_mg},\\
\frac{\partial \rho \epsilon}{\partial t} + \nabla\cdot\left[ \rho \epsilon \vb + \rho \epsilon \left(1 - \epsilon \right) \deltav \right] & = & 0\label{eq:cons_md}.
\end{eqnarray}
Interestingly, using the theorem of transport with the velocity $\mathbf{v}$ instead of the gas and the dust velocities gives:
\begin{eqnarray}
\frac{\mathrm{d} M_{\rm g}}{\mathrm{d} t} & = & \int_{S} \rho \epsilon \left(1 - \epsilon \right) \deltav\cdot\mathrm{n} \mathrm{d}S,  \label{eq:int_mg_bary}\\
\frac{\mathrm{d} M_{\rm d}}{\mathrm{d} t} & = & - \int_{S} \rho \epsilon \left(1 - \epsilon \right) \deltav\cdot\mathrm{n} \mathrm{d}S .  \label{eq:int_md_bary}
\end{eqnarray}
In the case where the volume $V$ is the entire space, the surface integrals of Eqs.~\ref{eq:int_mg_bary} -- \ref{eq:int_md_bary} go to zero and $ \frac{\mathrm{d} M_{\rm g}}{\mathrm{d} t}  = \frac{\mathrm{d} M_{\rm d}}{\mathrm{d} t} = 0$. However, the gas and the dust masses are in general \textit{not} conserved for any given volume $V$. Physically, this comes from the fact that the advection velocity of $V$ differs from the specific advection velocities of each phase taken individually. The specific flux of gas density going outside the volume $V$ and the specific flux of dust density going inside $V$ are given by $\pm \rho \epsilon \left(1 - \epsilon \right) \deltav$ respectively and hence counterbalance each other. The advection of the mixture at the barycentric velocity $\vb$ preserves the \textit{total} mass $M$ of the volume given by
\begin{equation}
M \equiv \int \rho \mathrm{d} V =  \int \left( \rhog + \rhod \right) \mathrm{d} V ,
\end{equation}
since summing Eqs.~\ref{eq:int_mg_bary} and \ref{eq:int_md_bary} gives
\begin{equation}
\frac{\mathrm{d} M }{\mathrm{d}t} = 0 .
\end{equation}
Equivalently, this gives the local equation of conservation:
\begin{equation}
\frac{\partial \rho}{\partial t} + \nabla\cdot\left( \rho \vb \right)  =  0 ,
\label{eq:mass_cons}
\end{equation}
which can also be directly derived from the sum of Eqs.~\ref{eq:cons_mg} and \ref{eq:cons_md}. 

\subsubsection{Conservation of momentum}
The specific momentum of the gas and dust phases may be defined according to
\begin{eqnarray}
\mathbf{P}_{\rm g} & \equiv & \int_{V} \rhog \vg \mathrm{d} V =  \int_{V} \left(1 - \epsilon \right)\rho \left(\vb - \epsilon \deltav \right) \mathrm{d} V, \\
\mathbf{P}_{\rm d} & \equiv & \int_{V} \rhod \vd \mathrm{d} V =  \int_{V} \epsilon \rho \left( \vb + \left(1 - \epsilon  \right)\deltav \right) \mathrm{d} V,
\end{eqnarray}
where the forces should balance over the volume $V$. Neglecting external forces acting on the mixture and assuming the only surface forces are the gas and dust pressure gradients, we have
\begin{eqnarray}
\frac{\mathrm{d}_{\rm g} \mathbf{P}_{\rm g}}{\mathrm{d} t} &  \equiv & -\int_{S} P_{\rm g} \mathbf{n}\mathrm{d}S  , \label{eq:int_pg} \\
\frac{\mathrm{d}_{\rm d} \mathbf{P}_{\rm d}}{\mathrm{d} t} &  \equiv & 0  , \label{eq:int_pd}
\end{eqnarray}
where $P_{\rm g}$ is the gas pressure.
These in turn result in the local conservation equations:
\begin{eqnarray}
\frac{\partial \rho \left( 1 - \epsilon \right)\left(\vb  - \epsilon \deltav \right)}{\partial t} \nonumber \\ 
 + \nabla\cdot\left[ \rho \left( 1 - \epsilon\right) \left(\vb  - \epsilon \deltav \right) \left(\vb  - \epsilon \deltav \right) + P_{\rm g} \mathrm{\mathbf{I}}  \right]    &=& 0 \label{eq:cons_pg}. \\
\frac{\partial \rho\epsilon\left(\vb + \left(1 - \epsilon \right)\deltav \right)}{\partial t}  \nonumber \\
 + \nabla\cdot\left[ \rho \epsilon \left(\vb + \left(1 - \epsilon \right)\deltav \right) \left(\vb + \left(1 - \epsilon \right)\deltav \right)  \right]    &=& 0 \label{eq:cons_pd}.
\end{eqnarray}
If the volume is advected with the velocity $\mathbf{v}$, Eqs.~\ref{eq:int_pg} -- \ref{eq:int_pd} can be rewritten
\begin{eqnarray}
\frac{\mathrm{d} \mathbf{P}_{\rm g}}{\mathrm{d} t} &  = &  -\int_{S} P_{\rm g} \mathbf{n} \mathrm{d}S \nonumber \\
&& +  \int_{S} \rho \epsilon \left(1 - \epsilon \right) \left(\vb - \epsilon \deltav \right) \deltav \cdot \mathbf{n} \mathrm{d}S , \label{eq:int_pg_bary} \\
\frac{\mathrm{d} \mathbf{P}_{\rm d}}{\mathrm{d} t} &  = & -  \int_{S} \rho \epsilon \left(1 - \epsilon \right) \left(\vb + \left(1 - \epsilon\right) \deltav \right) \deltav \cdot \mathbf{n} \mathrm{d}S\cdot\label{eq:int_pd_bary}
\end{eqnarray}
The new terms in the right-hand sides of Eq.~\ref{eq:int_pg_bary} -- \ref{eq:int_pd_bary} consist of momentum fluxes associated with the density fluxes related to the differential advection of each phase (Eqs.~\ref{eq:int_mg_bary} -- \ref{eq:int_md_bary}). It is important to note that even if the conservation of the total mass in a local volume is ensured, the total momentum $\mathbf{P} = \mathbf{P}_{\rm g} + \mathbf{P}_{\rm d}$ is \textit{not} conserved for the local volume. Indeed, summing Eqs.~\ref{eq:int_pg_bary} and \ref{eq:int_pg_bary}, we obtain:
\begin{equation}
\frac{\mathrm{d} \mathbf{P}}{\mathrm{d} t} = -\int_{S} P_{\rm g} \mathbf{n} \mathrm{d}S  -  \int_{S} \rho \epsilon \left(1 - \epsilon \right) \deltav \deltav\cdot\mathrm{n} \mathrm{d}S .
\label{eq:p_bary}
\end{equation}
Thus, even if mass fluxes of gas and dust counterbalance each other, they bring a net flux of total momentum in $V$ which is similar to the contribution of an anisotropic pressure. Using the gradient and the divergence theorems, Eq.~\ref{eq:p_bary} results in a local conservation equation given by
\begin{equation}
\frac{\partial \rho \vb}{\partial t} + \nabla\cdot\left[ \rho \vb \vb + P_{\rm g}  \mathrm{\mathbf{I}} + \rho \epsilon \left(1 - \epsilon \right) \deltav \deltav \right]  =  0,
\end{equation}
which can also be obtained by summing Eqs.~\ref{eq:int_pg_bary} and \ref{eq:int_pd_bary}. Note that this anisotropic pressure term brings additional terms in the energy equation as well (see below). In the case where $V$ is the entire space, the surface terms go to zero and $\frac{\mathrm{d}\mathbf{P}}{\mathrm{d}t} = 0$.

\subsubsection{Conservation of energy}
 The total energy of the mixture in a given volume $V$ is given by
\begin{eqnarray}
E &  \equiv &  \int_{V} \left( \frac{1}{2}\rhog \vg^{2} + \frac{1}{2}\rhod \vd^{2} + \rhog u      \right) {\rm d}V ,  \label{eq:olde} \\
& = &  \int_{V} \left( \frac{1}{2}\rho \vb^{2} + \frac{1}{2}\frac{\rhog \rhod}{\rho} \deltav^{2} + \left(1 - \epsilon \right)\rho u  \right) {\rm d}V  \label{eq:newe}.
\end{eqnarray}
Writing down the energy conservation for both the gas and the dust gives
\begin{eqnarray}
\frac{\mathrm{d}_{\rm g} E_{\rm g}}{\mathrm{d} t} & = & -\int_{S} P_{\rm g} \left(\vb - \epsilon \deltav \right) \mathbf{n} \mathrm{d}S, \\
\frac{\mathrm{d}_{\rm d} E_{\rm d}}{\mathrm{d} t} & = & 0 .
\end{eqnarray}
and thus two local conservation equations of the form
\begin{eqnarray}
\frac{\partial \left( \rho \left(1 - \epsilon \right) \left(u + \frac{1}{2}\left(\vb - \epsilon \deltav \right)^{2} \right) \right) }{\partial t} &&\\ \nonumber
+  \nabla \cdot  \left( \rho \left(1 - \epsilon \right) \left(u + \frac{1}{2}\left(\vb - \epsilon \deltav \right)^{2} + P_{\rm g}\right) \left(\vb - \epsilon \deltav \right) \right) & = & 0 , \label{eq:local_eg}\\
\frac{\displaystyle \partial \left( \frac{\rho \epsilon}{2} \left( \vb + \left( 1 - \epsilon\right) \deltav \right)^{2} \right) }{\partial t} &&\\
+  \nabla \cdot \left( \frac{\rho \epsilon}{2} \left( \vb + \left( 1 - \epsilon\right) \deltav  \right)^{2} \left(\vb + \left( 1 - \epsilon\right) \deltav \right) \right) & = & 0 \label{eq:local_ed}.
\end{eqnarray}
Combining Eqs.~\ref{eq:local_eg} and \ref{eq:local_ed} gives the total energy equation in conservative form:
\begin{eqnarray}
 \frac{\displaystyle \partial e }{\partial t} +  \nabla \cdot \Bigg\lbrace \left( \frac{1}{2}\rho \vb^{2} + \frac{1}{2}\rho \epsilon\left(1 - \epsilon \right) \deltav^{2} \right) \vb  \nonumber &&\\ 
 + \frac{\rho}{2}\left( 2\epsilon\left(1 - \epsilon \right)\vb \deltav + \epsilon\left(1 - \epsilon \right) \left(1 - 2 \epsilon \right) \deltav^{2} \right) \deltav \nonumber && \\ 
 + \rho\left(1 - \epsilon \right)\left(u + P_{\rm g} \right)\left(\vb - \epsilon \deltav \right) \Bigg\rbrace  &= &0 .
\label{eq:local_et}
\end{eqnarray}
where
\begin{equation}
e \equiv \frac{1}{2}\rho \vb^{2} + \frac{1}{2}\rho\epsilon\left(1 - \epsilon \right) \deltav^{2} + \left(1 - \epsilon \right)\rho u.
\end{equation}
As was the case for the momentum, $\frac{\mathrm{d}E}{\mathrm{d}t} = 0$ if $V$ is the entire space, since the surface terms tend to zero.

\subsubsection{One-fluid equations in conservative form}
Summarising, conservation laws of physical quantities result in local equations of evolution that can be written in a conservative form where only partial time derivatives of physical quantities and divergences of their fluxes are involved. Adding the remaining drag contribution as a source term, the equations of evolution of the mixture in conservative form are:
\begin{eqnarray}
\frac{\partial \rho}{\partial t} + \nabla\cdot\left( \rho \vb \right) & = & 0 \label{eq:cons_rho},\\
\frac{\partial \rho \epsilon}{\partial t} + \nabla\cdot\left[ \rho \epsilon \vb + \rho \epsilon \left(1 - \epsilon \right) \deltav \right] & = & 0,\label{eq:cons_epsd}\\
\frac{\partial \rho \vb}{\partial t} + \nabla\cdot\left[ \rho \vb \vb + P_{\rm g} \mathrm{\mathbf{I}} + \rho \epsilon \left(1 - \epsilon \right) \deltav \deltav \right] & = & 0,\\
\frac{\partial \rho\epsilon\left(\vb + \left(1 - \epsilon \right)\deltav \right)}{\partial t} \nonumber \\
 + \nabla\cdot\left[ \rho \epsilon \left(\vb + \left(1 - \epsilon \right)\deltav \right) \left(\vb + \left(1 - \epsilon \right)\deltav \right)  \right] &=& -K\deltav \label{eq:cons_vd},\\
 \frac{\displaystyle \partial e }{\partial t} +  \nabla \cdot \Bigg\lbrace \left( \frac{1}{2}\rho \vb^{2} + \frac{1}{2}\rho \epsilon\left(1 - \epsilon \right) \deltav^{2} \right) \vb  \nonumber &&\\ 
 + \frac{\rho}{2}\left( 2\epsilon\left(1 - \epsilon \right)\vb \deltav + \epsilon\left(1 - \epsilon \right) \left(1 - 2 \epsilon \right) \deltav^{2} \right) \deltav \nonumber && \\ 
 + \rho\left(1 - \epsilon \right)\left(u + P_{\rm g} \right)\left(\vb - \epsilon \deltav \right) \Bigg\rbrace  &= &0 .\label{eq:cons_en}
\end{eqnarray}
No drag term is involved in Eq.~\ref{eq:cons_en} since the \textit{total} energy of the mixture is rigorously conserved. Compared to the single fluid case, additional fluxes appear related to the differential advection between the total mass of the mixture and the masses. It is a simple matter of algebra to show that those equations are strictly equivalent to the equations where the quantities are advected with the velocity $\vb$. However, the conservative formulation is more relevant for Eulerian methods.

\subsection{Hyperbolicity}

Hyperbolic systems partial differential equation (PDE) are a particular category of PDE often encountered in hydrodynamics, for which specific powerful methods of resolution have been developed (Riemann solvers, e.g. \citealt{Toro1999}). Writing a PDE system over the variables $\mathbf{W} $ in the primitive form:
\begin{equation}
\partial_{t} \mathbf{W} + A \partial_{x} \mathbf{W} = 0 ,
\end{equation}
the system is hyperbolic if all the eigenvalues of the matrix $A$ are real. The hyperbolicity of a system of partial differential equations is related to the conservative nature of the equations. A seminal hyperbolic PDE system is formed by the mass and the momentum equation for an isothermal gas. Physically, a linear perturbation of this system provides a sound wave advected by the background flow. In a two-fluid model, the PDE system describing the evolution of a gas and dust mixture are trivially hyperbolic, since they consists of two independent systems of $2\times 2$ hyperbolic equations. The mathematical transformation to go from this two-fluid system to the one-fluid description involves i) a change of coordinate which does not involves the derivatives of the physical quantities, performed by applying the related Jacobian matrix and ii) a linear change in the advection velocities. Those are both linear transformations. Thus, the eigenvalues of the matrix $\mathbf{\mathrm{A}}$ are all real as the eigenvalues of a similar matrix are real and the hyperbolicity of the system is preserved. In a one dimensional case with a two fluids description, $\mathbf{W} = \left(\rho_{\rm g}, v_{\rm g} , \rho_{\rm d} , v_{\rm d}  \right)$ and the matrix $A$ is given by:
\begin{equation}
A = 
\begin{pmatrix}
v_{\rm g} & \rho_{\rm g} & 0 & 0 \\
\frac{c_{\rm s}^{2}}{\rho_{\rm g}} & v_{\rm g} & 0 & 0 \\
0 & 0 &v_{\rm d}  & \rho_{\rm d}\\
0 & 0 & 0 & v_{\rm d} 
\end{pmatrix} ,
\end{equation}
whose four eigenvalues are $v_{\rm g} \pm c_{\rm s}, v_{\rm d}, v_{\rm d}$. If $\tilde{W} = \left(\rho,\epsilon,v,\Delta v \right)$ (one fluid description), the matrix $\tilde{A}$ is:
\begin{equation}
\tilde{A} = 
\begin{pmatrix}
v & 0 & \rho & 0 \\
\frac{ \epsilon \left(1 - \epsilon \right) \Delta v}{\rho} & v + \Delta v \left(1 - 2 \epsilon \right) & 0 & \epsilon \left(1 - \epsilon \right) \\
\frac{ \epsilon \left(1 - \epsilon \right) \Delta v ^{2}}{\rho} + \frac{\left(1 - \epsilon \right)c_{\rm s}^{2}}{\rho} & \left(1 - 2\epsilon \right)\Delta v^{2} - c_{\rm s}^{2} &v  & 2 \Delta v \epsilon \left(1 - \epsilon \right)\\
- \frac{c_{\rm s}^{2}}{\rho} & - \Delta v^{2} + \frac{c_{\rm s}^{2}}{1 - \epsilon} & \Delta v& v - \left(2\epsilon - 1 \right)\Delta v 
\end{pmatrix} .
\end{equation}
Its four real eigenvalues are $v -\epsilon \Delta v \pm c_{\rm s}, v + \left(1 - \epsilon \right)\Delta v,, v + \left(1 - \epsilon \right)\Delta v $, which are nothing else than the eigenvalues of the matrix $A$ in the new system of coordinates, as expected. This corroborates the fact that the equations of the mixture can also be written in a conservative form (Eqs.~\ref{eq:cons_rho} -- \ref{eq:cons_vd}). To solve the Riemann problem, the eigenvectors of $A$ the ones of the two-fluids problem in the new system of coordinate. With a Riemann solver, the drag is integrated as a simple source term of the system.

\section{Simplified equations for strong drag / small grains}
\label{sec:approx}

\subsection{Comparison of terms}
To analyse the respective order of magnitude of the different terms, we define two dimensionless quantities: the traditional Mach number $M$
\begin{equation}
M \equiv \frac{\vb^{2}}{\cs^{2}} ,
\label{eq:mach}
\end{equation}
and the differential Mach number
\begin{equation}
M_{\Delta} \equiv \frac{\deltav^{2}}{\cs^{2}}  .
\label{eq:diffmach}
\end{equation}
Using $T$ to denote the time it takes for the fluid to propagate a distance $L$  at a sound speed $\cs$, Eq.~\ref{eq:momentum_deltav} provides the ratio between $M$ and $M_{\Delta}$ as a function of $\ts / T$:
\begin{eqnarray}
\frac{M_{\Delta}}{M} & = & \mathcal{O}\left( 1 \right) \hspace{4.9mm} \mathrm{if} \hspace{1mm} \ts \gg T, \label{eq:smalldrag} \\
\frac{M_{\Delta}}{M} & =  & \mathcal{O}\left(\frac{\ts^{2}}{T^{2}} \right)  \hspace{2mm} \mathrm{if}  \hspace{1mm} \ts \ll T.  \label{eq:largedrag}
\end{eqnarray}
Eqs.~\ref{eq:smalldrag} and \ref{eq:largedrag} define the weak and strong drag regimes respectively.

\subsection{Terminal velocity approximation}
\label{sec:eqstrong}

As discussed in \citetalias{LP12a}, strong drag regimes are tremendously difficult to handle with two fluids. They can however be very easily treated by the one fluid model since in this limit the differential velocity adds only a small correction to the barycentric velocity of the mixture. From Eq.~\ref{eq:largedrag} 
\begin{equation}
\left\Vert \frac{ \partial_{t} \deltav }{\deltav / \ts} \right\Vert= \left\Vert \frac{\vb\cdot\nabla \deltav }{\deltav / \ts} \right\Vert=  \left\Vert \frac{  \deltav\cdot\nabla \vb }{\deltav / \ts} \right\Vert= \mathcal{O}\left(\frac{\ts}{T} \right) .
\label{eq:firstorder}
\end{equation}
Thus, for small grains, Eq.~\ref{eq:momentum_deltav} simply reduces to
\begin{equation}
\deltav = \frac{\nabla P_{\mathrm{g}}}{\rhog} \ts ,
\label{eq:term_velocity}
\end{equation}
which is known as the \emph{terminal velocity approximation} \citep{Youdin2005,Chiang2008,Barranco2009,Lee2010,Jacquet2011}. Indeed, $\deltav / \vb$ is of order $\ts / T$ and terms containing $\deltav ^{2}$ can safely be neglected since they are second order in $\ts/T$. In the general case of forces acting on each phase separately (Sec.~\ref{sec:gen}), the terminal velocity approximation corresponds to
\begin{equation}
\deltav = ({\bf f}_{\rm d} - {\bf f}_{\rm g}) \ts.
\label{eq:gen_term_velocity}
\end{equation}

\subsection{First order approximation}
\label{sec:first}
To first order in $\ts / T$, Eqs.~\ref{eq:mass_rho} -- \ref{eq:momentum_deltav} become
\begin{eqnarray}
\frac{\mathrm{d} \rho}{\mathrm{d} t} & = & -\rho (\nabla\cdot\vb) \label{eq:mass_single},\\
\frac{\mathrm{d} \vb}{\mathrm{d}t}  & = & \mathbf{f} - \frac{\nabla P_{\mathrm{g}}}{\rho}   \label{eq:momentum_single},\\[1em]
\frac{\mathrm{d}}{\mathrm{d}t} \left( \frac{\rhod}{\rhog} \right)  & = & -\frac{\rho}{\rhog^{2}} \nabla\cdot\left(\frac{\rhog \rhod}{\rho} \left[ \frac{\nabla P_{\mathrm{g}}}{\rhog} \ts  \right] \right) \label{eq:dtg_single}.
\end{eqnarray}
Equivalently, the dust fraction $\epsd$ can be used, giving
\begin{equation}
\frac{\mathrm{d}\epsd}{\mathrm{d}t} = -\frac{1}{\rho} \nabla\cdot\left(\frac{\rhog \rhod}{\rho} \left[ \frac{\nabla P_{\mathrm{g}}}{\rhog} \ts  \right] \right) \label{eq:eps_single},
\end{equation}
in place of Eq.~\ref{eq:dtg_single}.
Thus, in strong drag regimes, gas-dust mixtures can be described by the usual equations of fluid dynamics (with a modified sound speed, see below) coupled with one additional advection-diffusion equation for the dust-to-gas ratio. The relative drift from the dust with respect to the gas is taken into account by the term on the right hand side of Eq.~\ref{eq:dtg_single}. This source term results from the fact that dust tends to accumulate in the pressure maxima and, by conservation of momentum, pushes the gas outside. The term contains a spatial second derivative and thus acts to diffuse the dust-to-gas ratio, with the effective diffusion $\nu_{\mathrm{s}}$ being of order
\begin{equation}
\nu_{\mathrm{s}} \simeq \cs ^{2} \ts .
\label{eq:nus}
\end{equation}
This diffusion-like process is therefore slow compared to the propagation of the mixture and can be easily be integrated explicitly in a numerical scheme. It should be noted that if additional forces act on the gas phase only (e.g. viscous terms, magnetic forces), they have to be added in the terminal velocity approximation and thus, inside the diffusion term in Eq.~\ref{eq:dtg_single} as well.

The internal energy equation simplifies to
\begin{eqnarray}
\frac{{\rm d} u}{{\rm d} t} & = &  -\frac{P_{\mathrm{g}}}{\rhog} (\nabla\cdot\vb) -  \frac{P_{\mathrm{g}}}{\rhog} \nabla\cdot\left( \frac{\rhod}{\rho}\left[ \frac{\nabla P_{\mathrm{g}}}{\rhog} \ts  \right]  \right)  - \frac{\rhod}{\rhog} \frac{1}{\rho} \left(\frac{\nabla P_{\mathrm{g}}}{\rhog} \ts \cdot\nabla \right) \rho  \nonumber \\
&& + \frac{\rhod}{\rho} \left[ \frac{\nabla P_{\mathrm{g}}}{\rhog}  \right]^{2} \ts. \label{eq:newu_single}
\end{eqnarray}

 We show in Sec.~\ref{sec:dustywave} and \ref{sec:SI} that the reduced set of equations \ref{eq:mass_single}--\ref{eq:dtg_single} are sufficient to describe some of the most important physical effects of dust in astrophysics, including waves in a two fluid mixture and the streaming instability in protoplanetary discs.

In conservative form, Eqs.~\ref{eq:mass_single} -- \ref{eq:eps_single} are equivalent to:
\begin{eqnarray}
\frac{\partial \rho}{\partial t} + \nabla\cdot\left( \rho \vb \right) & = & 0 \label{eq:cons_rho_single},\\
\frac{\partial \rho \epsilon}{\partial t} + \nabla\cdot\left( \rho \epsilon \vb + \rho \epsilon \left(1 - \epsilon \right) \left[ \frac{\nabla P_{\mathrm{g}}}{\rhog} \ts  \right]  \right) & = & 0,\label{eq:cons_epsd_single}\\
\frac{\partial \rho \vb}{\partial t} + \nabla\cdot\left( \rho \vb \vb + P_{\rm g}  \mathrm{\mathbf{I}}  \right) & = & 0 \label{eq:cons_vd_single}. \\
\frac{\displaystyle \partial  \tilde{e} }{\partial t} +  \nabla \cdot \left\{ \left[ \frac{\rho \vb^{2}}{2} + \rho\left(1 - \epsilon \right)\left(u + P_{\rm g} \right)  \right] \vb  \right\} \nonumber \\
+ \nabla \cdot \left( \rho \left( \frac{\vb^{2}}{2} - \epsilon\left(1 - \epsilon \right) \left( u + P_{\rm g} \right) \right)\left[ \frac{\nabla P_{\mathrm{g}}}{\rhog} \ts  \right] \right)  & = &0. \label{eq:cons_u_single}
\end{eqnarray}
where
\begin{equation}
\tilde{e} \equiv \frac{1}{2}\rho \vb^{2} + \left(1 - \epsilon \right)\rho u.
\end{equation}

\subsection{Zeroth order approximation}
\label{sec:zeroth}

To zeroth order in $\ts / T$, $\ts = 0$ (i.e., the limit of infinite drag/perfect coupling), Eqs.~\ref{eq:mass_rho} -- \ref{eq:momentum_deltav} simply reduce to
\begin{eqnarray}
\frac{\mathrm{d} \rho}{\mathrm{d} t} & = & -\rho (\nabla\cdot\vb) \label{eq:mass_zero},\\
\frac{\mathrm{d} \vb}{\mathrm{d}t}  & = & \mathbf{f} - \frac{\nabla P_{\mathrm{g}}}{\rho} ,  \label{eq:momentum_zero} \\
\frac{\mathrm{d}}{\mathrm{d}t} \left( \frac{\rhod}{\rhog} \right)  & = & 0 \label{eq:dtg_zero},
\end{eqnarray}
with $\deltav = 0$. Eq.~\ref{eq:newu} reduces to
\begin{equation}
\frac{{\rm d} u}{{\rm d} t} =  -\frac{P_{\mathrm{g}}}{\rhog} (\nabla\cdot\vb). \label{eq:newu_zero}
\end{equation}
Using the dust fraction in place of the dust-to-gas ratio, we find
\begin{equation}
\frac{\mathrm{d}\epsd}{\mathrm{d}t} = 0,
\end{equation}
as an alternative to Eq.~\ref{eq:dtg_zero}.

 From Eq.~\ref{eq:momentum_single}, we see that the sources of momentum are the same as those for the perfect gas, except that the gas pressure gradient pushes the whole mixture (and not the gas only). Since
\begin{equation}
\frac{\nabla P_{\mathrm{g}}}{\rho} = \frac{1}{\left( 1 + \rhod / \rhog \right)}\frac{\nabla P_{\mathrm{g}}}{\rhog} ,
\label{eq:grappzero}
\end{equation}
a dust and gas mixture can be treated to zeroth order as a single fluid whose sound speed $\tilde{c}_{\rm s}$ is the gas sound speed corrected by a factor depending on the dust-to-gas ratio
\begin{equation}
\tilde{c}_{\rm s} = c_{\rm s} \left(1 + \frac{\rhod}{\rhog}\right)^{-\frac12} .
\label{eq:cstilde}
\end{equation}
 This zeroth order effect is the main piece of physics required to correctly simulate the propagation of shocks in compressible dust-gas mixtures, for example in the interstellar medium (see Sec.~\ref{sec:dustyshock}).
In conservative form, Eqs.~\ref{eq:mass_zero} -- \ref{eq:dtg_zero} are equivalent to:
\begin{eqnarray}
\frac{\partial \rho}{\partial t} + \nabla\cdot\left( \rho \vb \right) & = & 0 \label{eq:cons_rho_zero},\\
\frac{\partial \rho \epsilon}{\partial t} + \nabla\cdot\left( \rho \epsilon \vb  \right) & = & 0,\label{eq:cons_epsd_zero}\\
\frac{\partial \rho \vb}{\partial t} + \nabla\cdot\left( \rho \vb \vb + P_{\rm g}  \mathrm{\mathbf{I}}  \right) & = & 0 .\label{eq:cons_vd_zero} \\
\frac{\displaystyle \partial \tilde{e} }{\partial t} +  \nabla \cdot \left( \left[ \frac{\rho \vb^{2}}{2} + \rho\left(1 - \epsilon \right)\left(u + P_{\rm g} \right)  \right] \vb  \right) & = &0. \label{eq:cons_u_zero}
\end{eqnarray}
%

\section{Applications}
\label{sec:tests}
Since the one fluid model given by Eqs.~\ref{eq:mass_rho} -- \ref{eq:momentum_deltav} is completely general, all of the analytic solutions described in \citet{LP11} and \citetalias{LP12a} can be equally well captured with the barycentric formulation. Here we show that not only are the solutions much simpler in the framework of our one fluid formulation, but also that most of the important physics can be captured with the reduced sets of equations presented in Secs.~\ref{sec:first} and \ref{sec:zeroth}. Thus, for many problems we expect that the zeroth or first order approximations will be sufficient.

\subsection{\textsc{dustybox}}
The \textsc{dustybox} problem \citep{Monaghan1995,PM2006,Miniati2010,LP11} consists of the drag-induced decay of the differential motion two fluids coupled by a drag term, assuming uniform densities $\rhog$ and $\rhod$. In the barycentric framework, assuming constant $\rhog$, $\rhod$ and $\vb$, the problem simply reduces to
\begin{equation}
\frac{\mathrm{d} \deltav}{\mathrm{d}t} = -\frac{\deltav}{\ts} ,
\end{equation}
where in general $\ts$ can itself be a function of $\deltav$. Since $\deltav$ is specifically evolved in the one fluid model --- in contrast to the two fluid approach where it depends on both the gas and the dust --- the \textsc{dustybox} problem is straightforward.

\begin{figure}
   \centering
   \includegraphics[angle=0, width=\columnwidth]{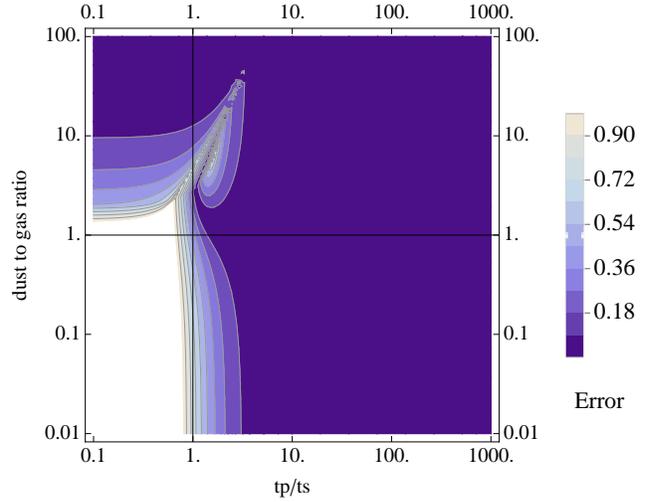} 
   \caption{Contour plot of the relative error in the dissipation timescale when using the terminal velocity approximation compared to the full \textsc{dustywave} problem, as a function of the ratio between the pressure time and the stopping time $ \tp / \ts$ (x axis) and the dust to gas ratio $\rhod / \rhog$ (y axis). The terminal velocity approximation is valid in the purple region (error less than $10\%$), especially if $ \tp / \ts$ is $\gtrsim$ 5-10 (error less than $1\%$). The approximation breaks down for small values of $ \tp / \ts$.}
   \label{fig:overdissip}
\end{figure}

\subsection{\textsc{dustywave} for strong drag}
\label{sec:dustywave}

The \textsc{dustywave} problem consists of linear sound waves propagating in one dimension in a dust and gas mixture of uniform density with a linear drag term. \textsc{dustywave} is solved in detail in \citet{LP11}, with the limit of strong drag regimes given in \citetalias{LP12a}. We assume that the equilibrium velocities and densities of the single fluid is given by $\vb = 0$, $\rho = \rhoz$ and $\rhod / \rhog = \rhodz / \rhogz$. We then consider small perturbations and perform an acoustic linear expansion of Eqs.~\ref{eq:mass_single}--\ref{eq:dtg_single}, i.e.
\begin{eqnarray}
  \frac{\partial \delta \rhog }{\partial t} +   \frac{\partial \delta \rhod }{\partial t} + \rhoz \frac{\partial \delta \vb }{\partial x}  & = & 0  ,\label{eq:lin1}\\ [1em]
  \rhoz \frac{\partial  \delta \vb }{\partial t} & = & - \cs^2 \frac{\partial \delta \rhog }{\partial x}  ,\label{eq:lin2} \\ [1em]
- \frac{\rhodz}{\rhogz^{2}}\frac{\partial \delta \rhog }{\partial t} + \frac{1}{\rhogz}  \frac{\partial \delta \rhod }{\partial t} & = & - \cs^{2} \ts \frac{\rhodz}{\rhogz^{2}}  \frac{\partial^{2} \delta \rhog }{\partial x^{2}} .\label{eq:lin3} 
\end{eqnarray}
As this system is linear, we search for solutions that have the form of monochromatic plane waves. The total solution is a linear combination of those monochromatic plane waves whose coefficients are fixed by the initial conditions. The perturbations are assumed to be of the general form
\begin{eqnarray}
\vb & = & V e^{i(kx - \omega t)} \label{eq:vpert}, \\
\delta\rhog & = & D_{\rm g} e^{i(kx - \omega t)} \label{eq:rhogpert}, \\
\delta\rhod & = & D_{\rm d} e^{i(kx - \omega t)}. \label{eq:rhodpert}
\end{eqnarray}
Using (\ref{eq:vpert})--(\ref{eq:rhodpert}) in (\ref{eq:lin1})--(\ref{eq:lin3}), and solving for non-trivial solutions gives the following dispersion relation:
\begin{equation}
w^{2} \left(1 + \frac{\rhodz}{\rhogz} \right) + i \omega \frac{\rhodz}{\rhogz} k^{2} \cs^{2} \ts - k^{2}\cs^{2} = 0,
\label{eq:disp_rel_single}
\end{equation}
which is, to first order in $\omega \ts$,
\begin{equation}
\omega = \pm k \tilde{\cs} - \frac{i}{2} \ts k^{2} \cs^{2} \frac{\rhodz}{\rhoz}.
\label{eq:disp_expand}
\end{equation}
This is precisely the solution found in \citetalias{LP12a} for the \textsc{dustywave} problem with two fluids in the strong drag regime. This illustrates that the diffusion-like term of Eq.~\ref{eq:dtg_single} contains the most important effect from the drift of the gas with respect to the dust. 

To test the validity of the terminal velocity approximation, we compute the relative over dissipation of the energy that occurs when using the diffusion-like term. Fig.~\ref{fig:overdissip} shows a contour plot of the relative error committed on the typical dissipation time of the terminal velocity approximation and the full \textsc{dustywave} problem as a function of the ratio between the pressure timescale $\tp = (k \cs)^{-1}$ (i.e the typical time for a sound wave to propagate across one wavelength) and the stopping time (x axis) and the dust to gas ratio $\rhod / \rhog$ (y axis). Those have been calculated as the inverse of the minimum of the imaginary part of the roots of the respective dispersion relations.

Fig.~\ref{fig:overdissip} shows that the terminal velocity approximation remains accurate so long as the stopping time ($\ts$) is smaller than the pressure timescale ($\tp$) by more than one order of magnitude. More precisely, for a dust to gas ratio of $10^{-2}$ typical of the interstellar medium, the over dissipation is $0.01 \%$, $1\%$, $4\%$, $20\%$, $70\%$ for $\tp / \ts = 100, 10, 5, 2, 1 $ respectively (at the transition of intermediate drag regimes, the approximation breaks quite abruptly). The best approximation is obtained when the dust to gas ratio is unity. The over dissipation is $0.04 \%$ for $\tp / \ts = 5$ and $\rhod / \rhog = 1$, and $15 \%$ for $\tp / \ts = 5$ and $\rhod / \rhog = 100$. Thus, we recommend keeping some safety margin by using the diffusion-like term only if the stopping time is smaller than the pressure time by an order of magnitude (or control it by a criteria similar to the one adopted in Fig.~\ref{fig:overdissip}).

In a numerical simulation, around 8 resolution lengths $\Delta x$ per wavelength is around the minimum resolution required to resolve a sound wave properly. Under this condition and using $\Delta t = \Delta x / \cs$ to denote the usual Courant time step, the condition for the diffusion approximation to remain valid becomes
\begin{equation}
\ts \lesssim \frac{\Delta t}{2\pi} .
\label{eq:crit}
\end{equation}
Thus, use of the terminal velocity approximation is appropriate roughly when the stopping time is comfortably shorter than the minimum timestep.

\subsection{Dustyshock}
\label{sec:dustyshock}

\begin{figure}
   \centering
   \includegraphics[angle=0, width=\columnwidth]{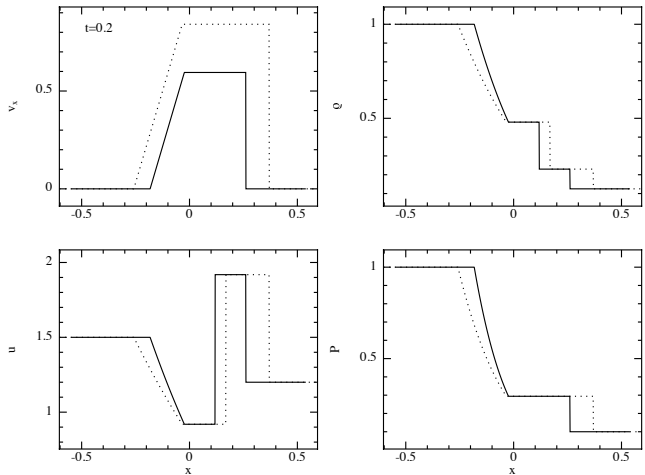} 
   \caption{Asymptotic solution of the \textsc{dustyshock} problem \citep{LP12a} for a dust-gas mixture with a dust to gas ratio of unity (solid lines), compared to the solution for a gas-only fluid (dotted lines), showing solutions for (clockwise) velocity, density, pressure and internal energy. The speed of the shock is strongly affected by the presence of dust (top left panel), showing that standard gas-only shock solutions are incorrect when the dust to gas ratio is high. In contrast to two fluid codes that require prohibitively high spatial resolution to capture the stationary phase of the \textsc{dustyshock} solution accurately, the solution can be captured trivially using our one fluid approach in the zeroth order approximation (Sec.~\ref{sec:zeroth}).}
   \label{fig:dustyshock}
\end{figure}

Shocks in dust-gas mixtures propagate in two distinct phases: a transient phase that occurs over a few stopping times during which the differential velocity between the gas and the dust is damped, and a stationary phase where the shock propagation is similar to that in a single fluid but with the modified sound speed $\tilde{\cs}$ defined by Eq.~\ref{eq:cstilde} (\citealt{Miura1982}; \citetalias{LP12a}). Thus, while capturing the transient phase requires solving the general one-fluid equations (i.e., Eqs.~\ref{eq:mass_rho}--\ref{eq:newu}), the essential physics during the stationary phase is captured by the zeroth order approximation discussed in Sec.~\ref{sec:zeroth}.

Fig.~\ref{fig:dustyshock} shows the exact solution to the \citet{Sod1978} shock tube problem in a gas (dotted lines), compared to the stationary solution for a dust-gas mixture (solid lines), employing the zeroth order approximation presented in Sec.~\ref{sec:zeroth} (here, the \textsc{dustyshock} solution was computed by simply modifying the sound speed in the shock tube exact solution distributed with \textsc{splash}; \citealt{Price2007}).

 Almost all astrophysical dusty shocks involve small grains (i.e. strong drag regimes), meaning that they are effectively in the stationary regime. Computing accurate solutions in this regime requires prohibitively high spatial resolution with two fluid codes ($\Delta x \lesssim c_{\rm s} t_{\rm s}$), as demonstrated by \citetalias{LP12a}. By contrast, the modifications required to solve Eqs.~\ref{eq:mass_zero}--\ref{eq:dtg_zero} with standard gas dynamics codes are essentially trivial, and the resolution requirements are identical to those in single-fluid codes, allowing an accurate treatment of the most important effects in shocks in regions where small dust grains are highly concentrated.

\subsection{Streaming instability for strong drag}
\label{sec:SI}

The streaming instability develops in dust and gas mixtures in differential rotation when the gas is submitted to an external background pressure gradient. In equilibrium, this background pressure gradient generates a differential velocity between the gas and the dust, leading the dust to migrate inwards and the gas outwards. The analytic expression for the radial and azimuthal velocities in both phases at stationary equilibrium have been derived by \citet{NSH86} (\citetalias{NSH86}). Perturbing this equilibrium can eventually lead to instabilities. In this case, energy is pumped from the background differential rotation due to the global pressure gradient and to force a perturbation in dust density to grow. This instability is particularly relevant for planet formation, since it provides a mechanism to accumulate enough solid material to form planetesimals locally. For an exhaustive survey of the streaming instability, we refer to \citet{Youdin2005,Youdin2007,Johansen2007,Jacquet2011}.

We will now prove that the set of equations Eqs.~\ref{eq:mass_single}  -- \ref{eq:dtg_single} is sufficient to recover the growth of the streaming instability in strong drag regimes. For transparency, we adopt the notations of \citet{Jacquet2011} for this section. We therefore introduce $\tstop$ defined by
\begin{equation}
\frac{\nabla P_{\mathrm{g}}}{\rhog} \ts \equiv \frac{\nabla P_{\mathrm{g}}}{\rho} \tstop ,
\label{eq:tstop}
\end{equation}
and consider for this problem that $\tstop$ is constant (though in general it depends on $\rhog$ and $\rhod$). For simplicity, we also assume that the gas is incompressible (such that $\delta \rhod = \delta \rho$) since the effect of compressibility on the streaming instability growth rate is negligible \citep{Youdin2005}. Given the fact that in the streaming instability problem, $\mathbf{f} = - \Omega^{2}(r) \mathbf{R} $, Eqs.~\ref{eq:mass_single} -- \ref{eq:dtg_single} straightforwardly provide the \citetalias{NSH86} solutions in the strong drag regime, i.e.
\begin{eqnarray}
\rhog & = & \rhogz \label{eq:rhog_zero},\\
\rhod & = & \rhodz \label{eq:rhod_zero},\\
\vbz & = &  \sqrt{ \left( r \Omega  \right)^{2} + r \ge} \, \etheta \label{eq:vb_zero},
\end{eqnarray}
where
\begin{equation}
\ge \equiv  - \left. \frac{\nabla P_{\mathrm{g}}}{\rho} \right|_{0} ,
\label{eq:defge}
\end{equation}
is a constant to be consistent with \citet{Jacquet2011}. Both gas and dust velocities are therefore entirely determined by adding Eq.~\ref{eq:term_velocity}, the terminal velocity approximation. Expanding Eqs.~\ref{eq:momentum_single} -- \ref{eq:dtg_single} to first order in a cartesian shearing box (i.e. neglecting the second-order curvature terms), we obtain:
\begin{eqnarray}
 \frac{\partial \delta \rho}{\partial t} + \frac{\partial \delta v_{r} }{\partial r} +  \frac{\partial \delta v_{z} }{\partial z}  & = & 0,\label{eq:rhoSI}\\
  \frac{\partial \delta v_{r} }{\partial t} + 2\Omega \delta v_{\theta} & = &  -\frac{1}{\rhoz} \frac{\partial \delta P_{\mathrm{g}}}{\partial r} - \ge \frac{\delta \rho}{\rho}    , \label{eq:vrSI} \\
  \frac{\partial \delta v_{\theta} }{\partial t} + \delta v_{r}\left( \Omega + \frac{\partial r\Omega}{\partial r}  \right)& = & 0, \label{eq:vthetaSI} \\
  \frac{\partial \delta v_{z} }{\partial t} & = & -\frac{1}{\rhoz} \frac{\partial \delta P_{\mathrm{g}}}{\partial z}, \label{eq:vzSI} \\
  \frac{1}{\rhogz} \frac{\partial \delta \rho }{\partial t} & = & - \frac{\rhoz}{\rhog} \tstop \left\lbrace    \frac{\rhodz}{\rhoz^{2}} \nabla \cdot\nabla\left(  \delta P_{\mathrm{g}} \right) \right. \\ \nonumber
  && \left. -\ge \mathbf{e}_{x}\frac{\rhoz^{2} \nabla\left( \delta \rho \right)  - \rhod \nabla\left( \delta \rho^{2}\right)  }{\rhoz^{3}} \right\rbrace.\label{eq:dtgSI}
\end{eqnarray}
Searching for non-trivial solutions of the form $e^{i(k_{x}x + k_{z}z - \omega t)}$, with $\kappa$ the epicyclic frequency and $k^{2} = k_{x}^{2} + k_{z}^{2}$, we obtain the following dispersion relation:
\begin{eqnarray}
-i\frac{\rhodz}{\rhoz}\tstop \omega^{4} + \omega^{3} + \left( i \frac{\rhodz}{\rhoz}\kappa^{2} + k_{x}\ge\frac{\rhogz}{\rhoz}    \right)\tstop \omega^{2} - \left(\kappa \frac{k_{z}}{k} \right)^2\omega && \\ \nonumber
+ k_{x}  \left(\kappa \frac{k_{z}}{k} \right)^2 \ge \tstop \frac{\rhodz - \rhogz}{\rhoz}    & = & 0, \label{eq:dispSI}
\end{eqnarray}
 which is identical to the one given by \citet{Jacquet2011}. This result is not surprising since the terminal velocity approximation is a good approximation for the streaming instability \citep{Youdin2005}. Importantly, this shows that that the physical processes relevant in protoplanetary discs are accounted for by the diffusion-like term in Eq.~\ref{eq:dtg_single}.

\section{Conclusion}
\label{sec:conclu}

We have shown how the two fluid equations describing the evolution of a mixture of dust and gas can be reformulated in terms of a single fluid moving with the barycentric velocity of the mixture. The formulation consists of differential equations for the total mass $\rho$, the barycentric velocity $\vb$, the differential velocity $\deltav$, and the dust to gas ratio $\rhod/\rhog$ (or equivalently, the dust fraction $\rhod/\rho$) that can be written in a form appropriate for both Lagrangian and Eulerian codes. The first two of these are identical (for $\rho$) or only slight modifications (for $\vb$) of the usual equations of gas dynamics. Evolving $\deltav$ greatly simplifies the drag between the fluids, reducing it to a simple exponential decay, meaning that it is easy to solve this equation for both weak and strong drag regimes, and thus capture the dynamics of both small and large grains within the same formulation. Finally, explicit evolution of the dust to gas ratio means that it is straightforward to follow the concentration of solid material, which plays a crucial role in planet formation.

The one fluid approach solves with physics the two most fundamental issues related to numerical simulations of two fluid dust and gas mixtures. Firstly, the presence of only one resolution scale in the simulation means that the problem of over-concentration of one fluid below the resolution of the other cannot occur. Secondly, the equations reduce identically to single fluid gas dynamics in the limit of infinite drag, avoiding the need for both infinite spatial resolution and infinitesimally small timesteps that would be necessary with the two fluid approach.

We have also shown that strong drag regimes can be handled in an even simpler manner by adding a diffusion-like term in the equation governing the dust to gas ratio evolution. This approach was shown to capture most of the interesting physical processes in the mixtures, including the propagation of sound waves and shocks, and the linear growth of the streaming instability. This simplified formulation --- which can be implemented with only minor modifications to existing single fluid codes --- was shown to remain sufficiently accurate for use in numerical simulations provided the stopping time $\ts$ is smaller than the minimum (Courant) timestep.

The nature of the evolution equations mean that this formalism lends itself readily to implementation in existing numerical codes. In a companion paper \citep{LP14b}, we describe an implementation of the one-fluid formulation with the Smoothed Particle Hydrodynamics (SPH) method, though we stress that the approach is equally useful for both SPH and grid-based codes.

\section*{Acknowledgments}
We thank Ben Ayliffe, Sarah Maddison and Mark Hutchison for useful discussions. We also thank the anonymous referee for comments which have improved this paper. This project was funded by the Australian Research Council (ARC) Discovery project grant DP1094585. DJP is very grateful for funding via an ARC Future Fellowship, FT130100034.

\bibliography{dustSPH}

\begin{appendix}
\section{finite volume of dust particles}
\label{sec:theta}

If dust grains occupy a finite volume, the volume fraction $\theta$ defined by
\begin{equation}
\theta = 1 - \frac{ \hrhod }{ \rhod },
\label{eq:theta}
\end{equation}
is no longer equal to unity. In this case, the volume densities of gas and dust $\hrhog$ and $\hrhod$ are distinguished from the intrinsic densities denoted $\rhog$ and $\rhod$ according to
\begin{eqnarray}
\hrhod & = & (1 - \theta) \rhod, \\
\hrhog & = & \theta \rhog.
\end{eqnarray}
Eqs.~\ref{eq:mass_rho} -- \ref{eq:momentum_deltav} should be modified accordingly, i.e.
\begin{eqnarray}
\frac{\partial \hrho}{\partial t} + \nabla\left( \hrho \vb \right) & = & 0  \label{eq:gmass_rho},\\
\frac{\partial \vb}{\partial t} + (\vb\nabla) \vb & = & \mathbf{f} - \frac{\nabla\left( P_{\mathrm{g}} + P_{\mathrm{d}} \right) }{\hrho} - \frac{1}{\hrho}\nabla\cdot \left(\frac{\hrhog \hrhod}{\hrho} \deltav \deltav \right)   \label{eq:gmomentum_bary},\\[1em]
\frac{\partial}{\partial t} \left(\frac{\hrhod}{\hrhog} \right) + \vb\nabla \left(\frac{\hrhod}{\hrhog}\right) & = & -\frac{\hrho}{\hrhog^{2}} \nabla \cdot \left(\frac{\hrhog \hrhod}{\hrho} \deltav \right) \label{eq:gdtgevol} , \\
\frac{\partial \deltav}{\partial t} + (\vb \cdot \nabla) \deltav  & = &  - \frac{\deltav}{\ts} + \frac{\nabla P_{\mathrm{g}}}{\hrhog} \left[ \theta - \frac{1 - \theta}{\hrhod/ \hrhog}\right] - \frac{\nabla P_{\mathrm{d}}}{\hrhod}\nonumber \\
&& - (\deltav \cdot \nabla) \vb + \frac{1}{2}\nabla \left( \frac{\hrhod - \hrhog}{\hrhod + \hrhog} \deltav ^{2} \right)  \label{eq:gmomentum_deltav},
\end{eqnarray}
where $\hrho = \hrhog + \hrhod$ and $\vb = \displaystyle \frac{\hrhog \vg + \hrhod \vd}{\rhog + \rhod}$. We have also included a dust pressure $P_{\mathrm{d}}$ for complete generality.

\end{appendix}

\label{lastpage}
\end{document}